\documentclass[%
 reprint,
 superscriptaddress,
 amsmath,amssymb,
 aps,
 prl
]{revtex4-2}

\usepackage{graphicx}
\usepackage{dcolumn}
\usepackage{bm}
\usepackage{hyperref}
\usepackage{physics}
\usepackage[capitalise]{cleveref}

\newcommand{\cre}[1]{{b}^\dagger_{#1}}
\newcommand{\ann}[1]{{b}_{#1}}
\newcommand{\num}[1]{{n}_{#1}}

\begin{document}

\title{Anti-Drude Metal of Bosons}

\author{Guido Masella}
\affiliation{ISIS (UMR 7006) and icFRC, University of Strasbourg and CNRS, 67000 Strasbourg, France}
\author{Nikolay V. Prokof'ev}
\affiliation{Department  of  Physics,  University  of  Massachusetts,  Amherst,  Massachusetts  01003,  USA}
\affiliation{ISIS (UMR 7006) and icFRC, University of Strasbourg and CNRS, 67000 Strasbourg, France}
\author{Guido Pupillo}
\affiliation{ISIS (UMR 7006) and icFRC, University of Strasbourg and CNRS, 67000 Strasbourg, France}

\date{\today}

\begin{abstract}
    In the absence of frustration, interacting bosons in the ground state exist
    either in the superfluid or insulating phases. Superfluidity corresponds to
    frictionless flow of the matter field, and in optical conductivity is
    revealed through a distinct $\delta$-functional peak at zero frequency with
    the amplitude known as the Drude weight.  This characteristic low-frequency
    feature is instead absent in insulating phases, defined by zero static
    optical conductivity. Here we demonstrate that bosonic particles in
    disordered one dimensional, $d=1$, systems can also exist in a conducting,
    non-superfluid, phase when their hopping is of the dipolar type, often
    viewed as short-ranged in $d=1$. This phase is characterized by finite
    static optical conductivity, followed by a broad anti-Drude  peak at finite
    frequencies.  Off-diagonal correlations are also unconventional: they
    feature an integrable algebraic decay for arbitrarily large values of
    disorder.  These results do not fit the description of any known quantum
    phase and strongly suggest the existence of a novel conducting state of
    bosonic matter in the ground state.
\end{abstract}

\maketitle

Quantum phases of matter are distinguished by their static and dynamical
properties, quantified by  correlation functions.  For interacting bosonic
matter in one dimension, the superfluid phase is characterized by a
non-integrable algebraic decay of static one-body (off-diagonal) correlations
as a function of distance and by a $\delta$-functional peak at zero frequency
in the optical conductivity, respectively. The latter is reflecting a singular
response to a weak externally applied field.  Strong enough disorder can induce
a quantum phase transition from the superfluid to an insulating phase, known as
the Bose glass \cite{Giamarchi2003}.
In this phase, off-diagonal correlations decay exponentially with distance and
the optical conductivity starts from zero at zero-frequency, reflecting the
absence of long-lived collective modes at low-energy.  These two phases exhaust
the known possibilities for disordered bosons in one dimension in the absence
of frustration, where by frustration we understand a situation when the
path-integral representation of quantum statistics in imaginary time is not
sign-positive.  In this work, we provide numerical evidence for the existence
of a novel disorder-induced phase that is neither superfluid nor insulating.
Despite featuring an algebraic decay of off-diagonal correlations, it has zero
superfluid density and its optical conductivity is finite at zero frequency.
The latter is followed by a broad peak at a finite frequency of the order of
the nearest-neighbor hopping energy.  Because of this characteristic
"anti-Drude" behavior of optical conductivity, with finite minimum instead of
maximum at zero frequency, we term this novel phase an anti-Drude metal of
bosons (aDMB).

The aDMB phase is a result of interplay between interactions, disorder, and
particle hopping, which we choose to be of the dipolar type. The latter is
usually considered as short-ranged in $d=1$ \cite{Lahaye2009}. For
non-interacting models with short-range hopping, disorder is generally expected
to localize all wave-functions exponentially (Anderson localization)
\cite{Anderson1958}. However, recent theoretical works have demonstrated that
single particle states can localize algebraically in the presence of couplings
that decay with distance as a power-law
\cite{Botzung2019b,Deng2018,Nosov2019,deMoura2005,Celardo2016}.  What happens
in strongly interacting systems remained an open question, and this work
provides the first answers with the discovery of the aDMB ground state.

Dipolar couplings have been already experimentally realized for internal
excitations of cold magnetic atoms
\cite{dePaz2013,Baier2016,Lepoutre2019,Patscheider2020}, Rydberg excited atoms
\cite{Barredo2015,Orioli2018,Leseleuc2019}, ions
\cite{Richerme2014,Jurcevic2014}, and molecules \cite{Yan2013}.  The
propagation of excitations with dipolar couplings in the presence of disorder
is also highly relevant for a variety of solid-state systems, including nuclear
spins \cite{Alvarez2015}, nitrogen-vacancy centers in diamonds
\cite{Waldherr2014}, or two-level emitters placed near a photonic crystal
waveguide \cite{Hung2016}.

We note that the existence of a metallic bosonic phase has been suggested
previously \cite{Feigelman1993,Phillips2003,Motrunich2007}; e.g., in the
context of finite-temperature strange metal behavior of high-temperature
superconductors \cite{Phillips2003,Yang2019} and as a possible ground state in
lattice models with multi-particle interactions
\cite{Motrunich2007,Jiang2013,Block2011}. However, up to date, the existence of
a metallic phase of bosons has not been confirmed by exact methods in any
physical system. Since frustrated spin systems featuring a variety of
spin-liquids phases can be always re-formulated in terms of strongly
interacting bosons, we exclude frustrated models from this discussion.

We consider the following Hamiltonian for hard-core bosons confined to one
dimension
\begin{equation}\label{eq:hamiltonian}
    \mathcal{H} =
    -t \sum_{i < j} \frac{a^3}{\abs{r_{ij}}^3}
    \bqty{\cre{i}\ann{j} + \text{H.c.}}
    + \sum_i \epsilon_i \num{i} , \qquad (n_i \le 1).
\end{equation}
We employ standard notations for bosonic creation and annihilation operators on
site $i$ and occupation numbers, $\num{i} = \cre{i}\ann{i}$, that cannot exceed
unity in the allowed Fock states.  The nearest-neighbor hopping amplitude, $t$,
and the lattice spacing, $a$, are taken as units of energy and length,
respectively. Hopping amplitudes between sites $i$ and $j$ decay with the
distance between them as $r_{ij}^{-3}$, and $\epsilon_i$ are random on-site
energies uniformly distributed between $-W$ and $W$.  In spin language,
\cref{eq:hamiltonian} is equivalent to an XY Hamiltonian with dipolar
couplings, which, in the absence of disorder, can be realized in experiments
with cold polar molecules \cite{Yan2013}, trapped ions
\cite{Richerme2014,Jurcevic2014} and Rydberg atoms
\cite{Zeiher2017,Barredo2015,Orioli2018}, with the latter also in the presence
of disorder \cite{Leseleuc2019}.  Recent theoretical works provide strong
evidence that \cref{eq:hamiltonian} supports a many-body localized (MBL) phase
at finite energy \cite{Yan2013,Burin2015,Burin2015a,Safavi-Naini2019,Deng2020}.
Our result then implies that the MBL transition out of aDMB takes place as the
temperature is increased. In a system with an upper bound on the maximal energy
per particle this result is not that surprising \cite{Kagan1983}.

In the following, we determine the ground-state quantum phases of
\cref{eq:hamiltonian} using large scale path-integral quantum Monte-Carlo
simulations based on the Worm algorithm \cite{Prokofev1998}.  Without loss of
generality, we focus on particle density $\rho=1/2$.\\

\begin{figure}
    \centering
    \includegraphics[width=\columnwidth]{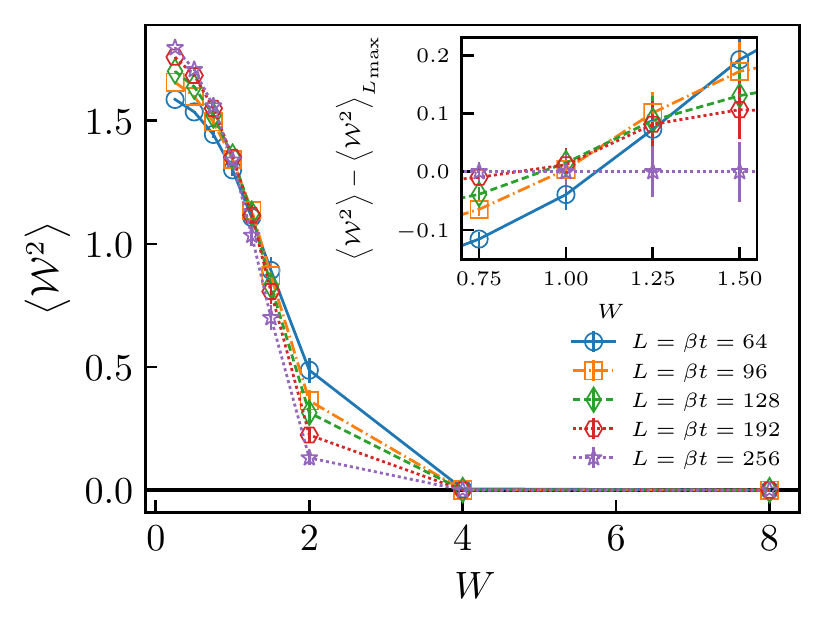}
    \caption{
        \label{fig:winding}
        Mean-squared winding numbers $\ev{\mathcal{W}^2}$ as functions of the
        disorder strength $W$ for lattice sizes $L=64$ (blue
        circles), $96$ (orange squares), $128$ (green diamonds), $192$ (red
        hexagons), $256$ (purple stars). Inset highlights the area near the
        phase transition, showing crossing points
        between the curves within the interval $W_c = 1.00 \pm 0.15$; the curve corresponding
        to the largest size ($L=256$) is subtracted from
        all data for clarity.
    }
\end{figure}

For nearest-neighbor hopping only, one-dimensional hard-core bosons behave as
spinless fermions and bosonic exchange has to involve all particles in the
liquid. A regular system would have finite superfluid density, $\rho_{\rm s}$,
that characterizes the response to twisted boundary condition caused by an
external vector potential field. It can be conveniently computed within quantum
Monte-Carlo,  see Methods, through the statistics of winding numbers,
$\mathcal{W}$, using the Pollock-Ceperley relation $\rho_{\rm s} \propto
\langle \mathcal{W} \rangle ^2$~\cite{Pollock1987}\footnote{See Supplementary
material}.  However, it is well known that the superfluid density of this
system is immediately suppressed by any finite strength of disorder $W$,  due
to Anderson localization~\cite{Giamarchi2003}.  Dipolar hopping changes this
picture entirely, by allowing for pair-wise bosonic exchanges, somewhat similar
to soft-core particles. One then expects superfluidity to be robust against
weak disorder, and, possibly, undergo a quantum phase transition to a
non-superfluid phase when disorder exceeds some critical value $W_{\rm c}$. \\

\Cref{fig:winding} shows numerical results for the statistics of mean-squared
winding numbers $\ev{\mathcal{W}^2}$ as a function of the disorder strength $W$
for different lattice sizes $L$. Mean-squared winding numbers
are expected to be scale invariant at a continuous phase transition, regardless
of the system dimension.  This allows one to identify the critical disorder
strength $W_c$ where superfluidity is lost
by the crossing point of the $\ev{\mathcal{W}^2}$-vs-$W$ curves for different
values of $L$.  The figure shows that all sizes larger than $L>64$ cross at
$W_{\rm c} = 1.00\pm 0.15$  (see Inset), signalling the transition from a
superfluid phase for $W<W_{\rm c}$ to a quantum phase that is not superfluid
for $W>W_{\rm c}$.  In the following, we focus on characterising the properties
of this non-superfluid phase with $W>W_{\rm c}$ by studying its correlation
functions and optical conductivity.\\

\begin{figure}
    \centering
    \includegraphics[width=\columnwidth]{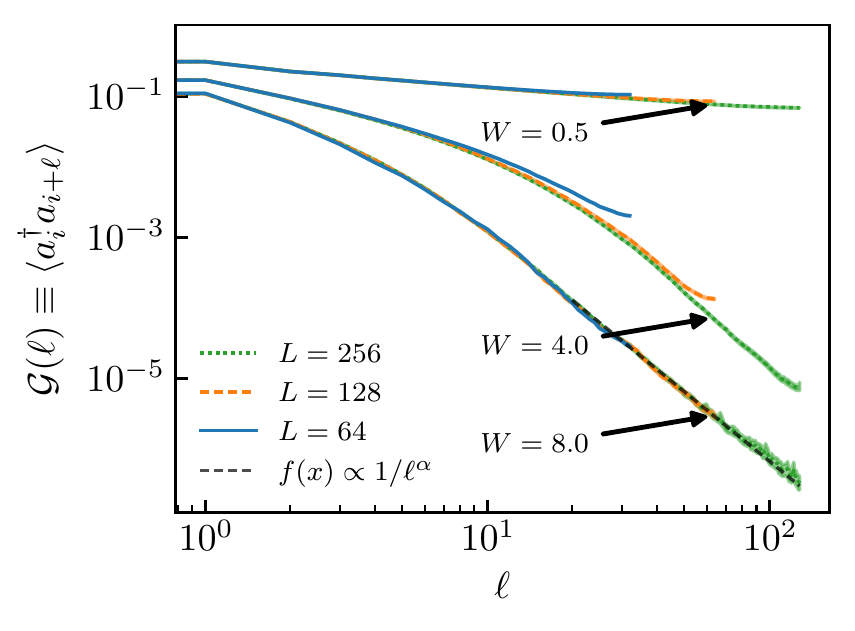}
    \caption{
        \label{fig:Green}
        One-body density matrix $\mathcal{G}(\ell)$ as a function of the
        distance $\ell$ for different system sizes $L=64$ (blue solid lines), $128$ (yellow dashed lines), and $256$ (green dotted lines), and values of the disorder strength $W = 0.5$, $4.0$, and $8.0$ (top to bottom). Data
        is shown on the doubly logarithmic scale. The gray dashed line corresponds to a fit $1/\ell^\alpha$ with $\alpha = 3.26(2)$ of the large distances
        behaviour for $L=256$ and $W=8.0$.
    }
\end{figure}

\begin{figure*}
    \centering
    \includegraphics[width=\textwidth]{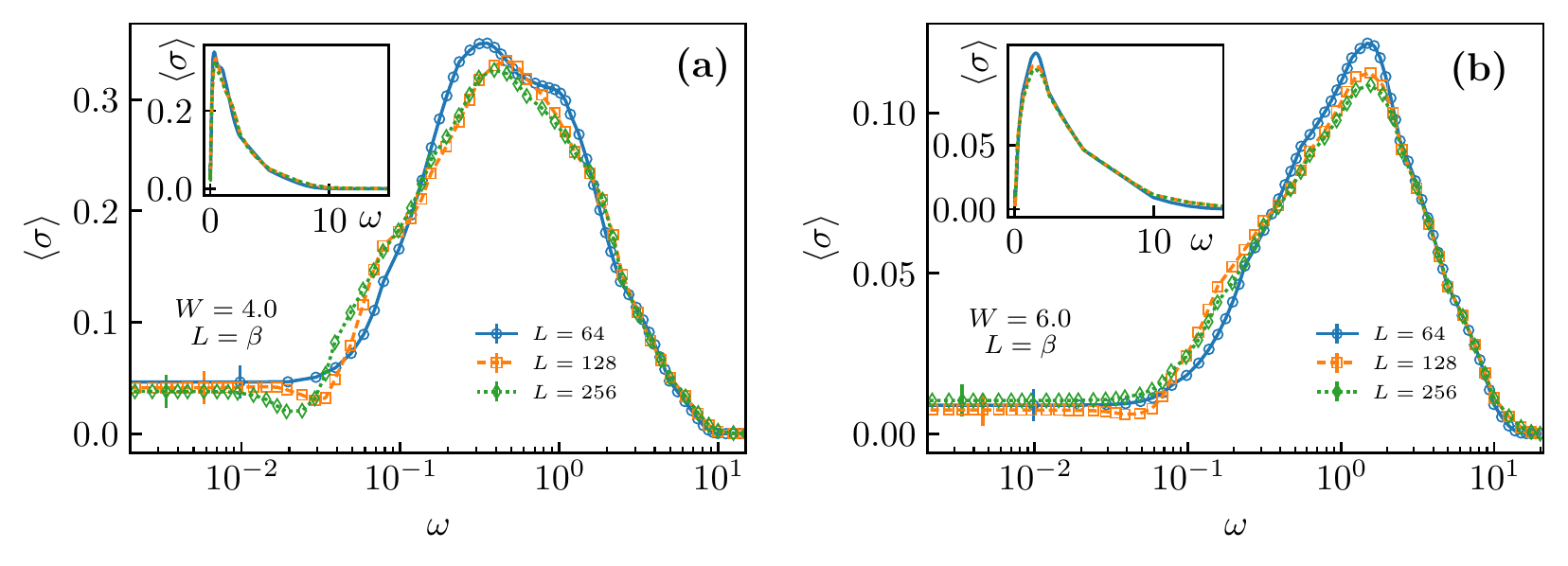}
    \caption{
        \label{fig:ConductivityAv}
        Disorder-averaged optical conductivity $\sigma$ as a function of the
        frequency $\omega$ for $W = 4$ (Panel a, left) and $W = 6$ (Panel b,
        right), at different system sizes $L=64$ (blue circles), $128$ (orange
        squares), and $256$ (green diamonds).  Data in the main plots is shown on the logarithmic scale for the frequency, highlighting the behaviour for
        small $\omega$. Insets show data on the linear scale.
    }
\end{figure*}

The one-body density matrix $\mathcal{G}(\ell)=\ev{\cre{i}\ann{i+\ell}}$ is
expected to decay algebraically as a function of distance $\ell$ for a
one-dimensional superfluid ground state, while in an insulating phase it is
expected to decay exponentially, e.g. in a crystalline phase or Bose glass.
\Cref{fig:Green} shows $\mathcal{G}(\ell)$ for the Hamiltonian
\cref{eq:hamiltonian}, for chosen values of the disorder strength $W$.  The
figure shows that in the superfluid phase with $W=0.5<W_{\rm c}$,
$\mathcal{G}(\ell)$ displays a slow algebraic decay, as expected.
Surprisingly, we find that an initial exponential decay of $\mathcal{G}(\ell)$
is followed at large distances $\ell$ by an algebraic decay in the
non-superfluid phase for $W>W_{\rm c}$. The large-distance decay is well
described by the $\mathcal{G}(\ell)\sim 1/\ell^3$ dependence. This behavior is
at odds with known results for insulating many-body phases with short-range
hopping \cite{Giamarchi2003}, indicating that other physical properties may
also be unconventional.  We thus proceed with analysing the optical
conductivity of the non-superfluid phase at $W>W_{\rm c}$.\\

The optical conductivity $\sigma(\omega)$ relates the current density $J$ to
the strength of an externally applied electric field $\mathcal{E}$ as
$J(\omega)=\sigma(\omega)\mathcal{E}(\omega)$, with $\omega$ the field
frequency. We obtain the optical conductivity $\sigma(\omega)$ within the
linear response theory by first computing the current-current correlation
function $\chi(\imath\omega_n) = \ev{j(\tau) j(0)}_{\imath\omega_n}$ at
Matsubara frequencies $\omega_n = 2\pi n T$ using the Worm algorithm, followed
by its numerical analytic continuation  (see Methods). Here $j$ is the lattice
current operator defined as $j = \imath t \sum_{i<j} r_{ij}
\bqty{\cre{i}\ann{j} - \cre{j}\ann{i}} / r_{ij}^3$.

\begin{figure*}
    \centering
    \includegraphics[width=\textwidth]{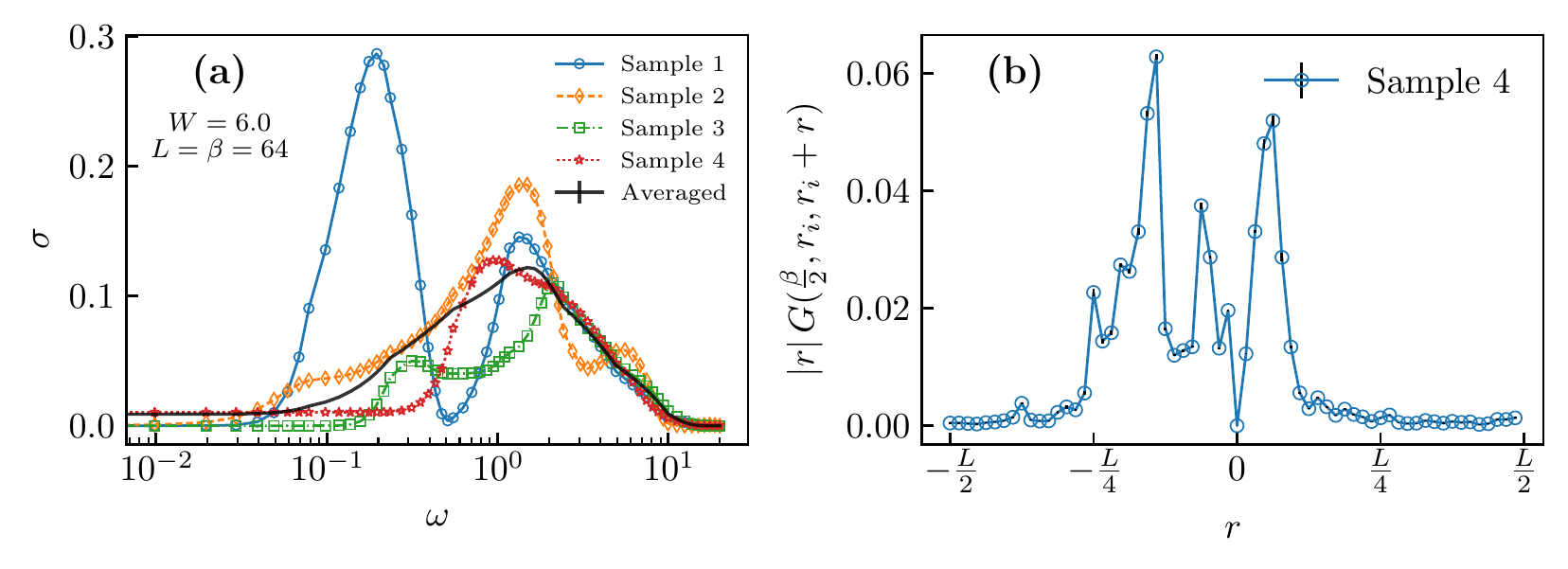}
    \caption{
        \label{fig:realizations}
        Panel a:
        Optical conductivity $\sigma$ as a function of the frequency $\omega$ for different disorder realizations. The black continuous line represents the average over all $384$ disordered samples.
        Panel b:
        Correlation function $|r|\,\mathcal{G}_{\tau =\frac{\beta}{2}}(r_\mathrm{i}, r_\mathrm{i} + r)$  as a function of $r$, sampled for imaginary time difference $\tau = \frac{\beta}{2}$ between the two end points on the trajectory.
        Here, $r_\mathrm{i}$ is chosen so that $\mathcal{G}_{\frac{\beta}{2}}(r_i, r_i)$ is maximum.
        This quantity allows one to visualize the main contributions to the current for a single disorder realization when the particle starts from point $r_i$ (see main text and Methods).
        In both panels  data is shown for $L = \beta = 64$ and $W =6$.
    }
\end{figure*}

\Cref{fig:ConductivityAv} shows typical examples of the optical conductivity,
averaged over a minimum of 384 disorder realizations, as a function of
frequency for two values of $W>W_{\rm c}$ deep in the non-superfluid phase and
different lattice sizes $L$. Consistently with the absence of superfluidity,
the figure shows that the characteristic  $\delta$-functional peak at zero
frequency peak in $\sigma(\omega\simeq 0)$ is absent.  However, the numerical
results also show two striking features: (i) The zero-frequency response is
finite and system size independent within the (relative large) error bars; (ii)
Unlike in usual conductors featuring a Drude peak (maximum at $\omega =0$), the
optical conductivity has a minimum at zero frequency followed by a large peak
at frequency $\omega \simeq t$, which provides a large response at energies of
the order of the nearest-neighbor hopping amplitude. This peak broadens with
increasing $W$, providing a large response up to frequencies $\omega\simeq 10
t$. These results for the averaged conductivity demonstrate the existence of a
conducting, non superfluid phase of bosons in the ground state.  This
conducting  behaviour is not due to well defined delocalized quasiparticle
states as in typical Drude-type metals; rather, it is an "anti-Drude metal" ,
where the largest response occurs at a small but finite frequency.\\

\Cref{fig:realizations}(a) shows selected results for $\sigma (\omega)$ in the
aDMB phase for individual realizations of disorder, i.e. without averaging. We
find that at frequencies $\omega > t$ the optical conductivity behavior is
rather robust and sample-to-sample fluctuations are not substantial.  The same
cannot be said about the low-frequency part that wildly fluctuates from sample
to sample - whilst some of the samples are metallic, the majority display an
insulating behavior.  This suggests that static $\sigma$ is in fact not a
self-averaging quantity in our system.  These fluctuations will be reflected in
similar fluctuations in experiments.\\

The discovery of the aDMB phase is particularly surprising as the dipolar
hopping term in Eq.~\eqref{eq:hamiltonian} is usually considered to be short
ranged in one dimension. Nevertheless, it leads to large de-localized
contributions to the current that can be visualized as follows. The single
particle propagator $ \mathcal{G}_{\tau}(r, r') =
\ev{\cre{r'}(\tau)\ann{r}(0)}$ encodes information for where a particle/hole
injected into the system at site $r$ can go in time $\tau$ (for hard core
bosons points $r$ and $r'$ are connected by a trajectory). By setting $\tau =
\beta / 2$ and taking the limit $\beta \to \infty$ we gain insight into
properties of the ground state wave function. Since current operator between
distant sites involves an additional power of distance we multiply $
\mathcal{G}_{\frac{\beta}{2}}(r, r')$ by $|r- r'|$ to establish a quantitative
measure for current contributions.  \Cref{fig:realizations}(b) visualizes the
correlation function $|r| \mathcal{G}_{\frac{\beta}{2}}(r_i, r_i +r)$ for a
single conducting realization as a function of the distance $r$ for a fixed
value of $r_i$ that was chosen from the condition of maximum for
$\mathcal{G}_{\frac{\beta}{2}}(r_i, r_i)$.  The figure makes it clear that
large current contributions are present over a wide range of distances of the
order of $\sim L/4$.\\

In summary, we have demonstrated that bosonic particles can exist in a unusual
metallic phase at zero temperature. It emerges from interplay between disorder,
interactions, and dipolar hopping that has already been realized in experiments
with Rydberg atoms, cold ions, and polar molecules. These results open many new
research directions.  These include investigations of new metallic phases that
can exist in higher dimensions and possible connections to the experimentally
observed ``bad metal" states on the finite-temperature phase diagram of
high-temperature superconductors.

\begin{acknowledgements}
    \emph{Acknowledgements} --
    The authors acknowledge support from the University of Strasbourg Institute
    of Advanced Studies (USIAS). G. P.  acknowledges additional support from
    the Institut Universitaire de France (IUF) and LABEX CSC.
    N. P.  acknowledges support from the MURI Program "New Quantum Phases of
    Matter" from AFOSR.
    Computing time was provided by the High Performance Computing Center of the
    University of Strasbourg. Part of the computing resources were funded by
    the Equipex Equip\@Meso project (Programme Investissements d'Avenir) and
    the CPER Alsacalcul/Big Data.
\end{acknowledgements}

\bibliography{library}

\begin{thebibliography}{41}%
\makeatletter
\providecommand \@ifxundefined [1]{%
 \@ifx{#1\undefined}
}%
\providecommand \@ifnum [1]{%
 \ifnum #1\expandafter \@firstoftwo
 \else \expandafter \@secondoftwo
 \fi
}%
\providecommand \@ifx [1]{%
 \ifx #1\expandafter \@firstoftwo
 \else \expandafter \@secondoftwo
 \fi
}%
\providecommand \natexlab [1]{#1}%
\providecommand \enquote  [1]{``#1''}%
\providecommand \bibnamefont  [1]{#1}%
\providecommand \bibfnamefont [1]{#1}%
\providecommand \citenamefont [1]{#1}%
\providecommand \href@noop [0]{\@secondoftwo}%
\providecommand \href [0]{\begingroup \@sanitize@url \@href}%
\providecommand \@href[1]{\@@startlink{#1}\@@href}%
\providecommand \@@href[1]{\endgroup#1\@@endlink}%
\providecommand \@sanitize@url [0]{\catcode `\\12\catcode `\$12\catcode
  `\&12\catcode `\#12\catcode `\^12\catcode `\_12\catcode `\%12\relax}%
\providecommand \@@startlink[1]{}%
\providecommand \@@endlink[0]{}%
\providecommand \url  [0]{\begingroup\@sanitize@url \@url }%
\providecommand \@url [1]{\endgroup\@href {#1}{\urlprefix }}%
\providecommand \urlprefix  [0]{URL }%
\providecommand \Eprint [0]{\href }%
\providecommand \doibase [0]{https://doi.org/}%
\providecommand \selectlanguage [0]{\@gobble}%
\providecommand \bibinfo  [0]{\@secondoftwo}%
\providecommand \bibfield  [0]{\@secondoftwo}%
\providecommand \translation [1]{[#1]}%
\providecommand \BibitemOpen [0]{}%
\providecommand \bibitemStop [0]{}%
\providecommand \bibitemNoStop [0]{.\EOS\space}%
\providecommand \EOS [0]{\spacefactor3000\relax}%
\providecommand \BibitemShut  [1]{\csname bibitem#1\endcsname}%
\let\auto@bib@innerbib\@empty
\bibitem [{\citenamefont {Giamarchi}(2003)}]{Giamarchi2003}%
  \BibitemOpen
  \bibfield  {author} {\bibinfo {author} {\bibfnamefont {T.}~\bibnamefont
  {Giamarchi}},\ }\href@noop {} {\emph {\bibinfo {title} {Quantum {{Physics}}
  in {{One Dimension}}}}}\ (\bibinfo  {publisher} {{Oxford University Press}},\
  \bibinfo {year} {2003})\BibitemShut {NoStop}%
\bibitem [{\citenamefont {Lahaye}\ \emph {et~al.}(2009)\citenamefont {Lahaye},
  \citenamefont {Menotti}, \citenamefont {Santos}, \citenamefont {Lewenstein},\
  and\ \citenamefont {Pfau}}]{Lahaye2009}%
  \BibitemOpen
  \bibfield  {author} {\bibinfo {author} {\bibfnamefont {T.}~\bibnamefont
  {Lahaye}}, \bibinfo {author} {\bibfnamefont {C.}~\bibnamefont {Menotti}},
  \bibinfo {author} {\bibfnamefont {L.}~\bibnamefont {Santos}}, \bibinfo
  {author} {\bibfnamefont {M.}~\bibnamefont {Lewenstein}},\ and\ \bibinfo
  {author} {\bibfnamefont {T.}~\bibnamefont {Pfau}},\ }\bibfield  {title}
  {\bibinfo {title} {The physics of dipolar bosonic quantum gases},\ }\href
  {https://doi.org/10.1088/0034-4885/72/12/126401} {\bibfield  {journal}
  {\bibinfo  {journal} {Reports on Progress in Physics}\ }\textbf {\bibinfo
  {volume} {72}},\ \bibinfo {pages} {126401} (\bibinfo {year}
  {2009})}\BibitemShut {NoStop}%
\bibitem [{\citenamefont {Anderson}(1958)}]{Anderson1958}%
  \BibitemOpen
  \bibfield  {author} {\bibinfo {author} {\bibfnamefont {P.~W.}\ \bibnamefont
  {Anderson}},\ }\bibfield  {title} {\bibinfo {title} {Absence of {{Diffusion}}
  in {{Certain Random Lattices}}},\ }\href
  {https://doi.org/10.1103/PhysRev.109.1492} {\bibfield  {journal} {\bibinfo
  {journal} {Physical Review}\ }\textbf {\bibinfo {volume} {109}},\ \bibinfo
  {pages} {1492} (\bibinfo {year} {1958})}\BibitemShut {NoStop}%
\bibitem [{\citenamefont {Botzung}\ \emph {et~al.}(2019)\citenamefont
  {Botzung}, \citenamefont {Vodola}, \citenamefont {Naldesi}, \citenamefont
  {M{\"u}ller}, \citenamefont {Ercolessi},\ and\ \citenamefont
  {Pupillo}}]{Botzung2019b}%
  \BibitemOpen
  \bibfield  {author} {\bibinfo {author} {\bibfnamefont {T.}~\bibnamefont
  {Botzung}}, \bibinfo {author} {\bibfnamefont {D.}~\bibnamefont {Vodola}},
  \bibinfo {author} {\bibfnamefont {P.}~\bibnamefont {Naldesi}}, \bibinfo
  {author} {\bibfnamefont {M.}~\bibnamefont {M{\"u}ller}}, \bibinfo {author}
  {\bibfnamefont {E.}~\bibnamefont {Ercolessi}},\ and\ \bibinfo {author}
  {\bibfnamefont {G.}~\bibnamefont {Pupillo}},\ }\bibfield  {title} {\bibinfo
  {title} {Algebraic localization from power-law couplings in disordered
  quantum wires},\ }\href {https://doi.org/10.1103/PhysRevB.100.155136}
  {\bibfield  {journal} {\bibinfo  {journal} {Physical Review B}\ }\textbf
  {\bibinfo {volume} {100}},\ \bibinfo {pages} {155136} (\bibinfo {year}
  {2019})}\BibitemShut {NoStop}%
\bibitem [{\citenamefont {Deng}\ \emph {et~al.}(2018)\citenamefont {Deng},
  \citenamefont {Kravtsov}, \citenamefont {Shlyapnikov},\ and\ \citenamefont
  {Santos}}]{Deng2018}%
  \BibitemOpen
  \bibfield  {author} {\bibinfo {author} {\bibfnamefont {X.}~\bibnamefont
  {Deng}}, \bibinfo {author} {\bibfnamefont {V.~E.}\ \bibnamefont {Kravtsov}},
  \bibinfo {author} {\bibfnamefont {G.~V.}\ \bibnamefont {Shlyapnikov}},\ and\
  \bibinfo {author} {\bibfnamefont {L.}~\bibnamefont {Santos}},\ }\bibfield
  {title} {\bibinfo {title} {Duality in {{Power}}-{{Law Localization}} in
  {{Disordered One}}-{{Dimensional Systems}}},\ }\href
  {https://doi.org/10.1103/PhysRevLett.120.110602} {\bibfield  {journal}
  {\bibinfo  {journal} {Physical Review Letters}\ }\textbf {\bibinfo {volume}
  {120}},\ \bibinfo {pages} {110602} (\bibinfo {year} {2018})}\BibitemShut
  {NoStop}%
\bibitem [{\citenamefont {Nosov}\ \emph {et~al.}(2019)\citenamefont {Nosov},
  \citenamefont {Khaymovich},\ and\ \citenamefont {Kravtsov}}]{Nosov2019}%
  \BibitemOpen
  \bibfield  {author} {\bibinfo {author} {\bibfnamefont {P.~A.}\ \bibnamefont
  {Nosov}}, \bibinfo {author} {\bibfnamefont {I.~M.}\ \bibnamefont
  {Khaymovich}},\ and\ \bibinfo {author} {\bibfnamefont {V.~E.}\ \bibnamefont
  {Kravtsov}},\ }\bibfield  {title} {\bibinfo {title} {Correlation-induced
  localization},\ }\href {https://doi.org/10.1103/PhysRevB.99.104203}
  {\bibfield  {journal} {\bibinfo  {journal} {Physical Review B}\ }\textbf
  {\bibinfo {volume} {99}},\ \bibinfo {pages} {104203} (\bibinfo {year}
  {2019})}\BibitemShut {NoStop}%
\bibitem [{\citenamefont {{de Moura}}\ \emph {et~al.}(2005)\citenamefont {{de
  Moura}}, \citenamefont {Malyshev}, \citenamefont {Lyra}, \citenamefont
  {Malyshev},\ and\ \citenamefont {{Dom{\'i}nguez-Adame}}}]{deMoura2005}%
  \BibitemOpen
  \bibfield  {author} {\bibinfo {author} {\bibfnamefont {F.~A. B.~F.}\
  \bibnamefont {{de Moura}}}, \bibinfo {author} {\bibfnamefont {A.~V.}\
  \bibnamefont {Malyshev}}, \bibinfo {author} {\bibfnamefont {M.~L.}\
  \bibnamefont {Lyra}}, \bibinfo {author} {\bibfnamefont {V.~A.}\ \bibnamefont
  {Malyshev}},\ and\ \bibinfo {author} {\bibfnamefont {F.}~\bibnamefont
  {{Dom{\'i}nguez-Adame}}},\ }\bibfield  {title} {\bibinfo {title}
  {Localization properties of a one-dimensional tight-binding model with
  nonrandom long-range intersite interactions},\ }\href
  {https://doi.org/10.1103/PhysRevB.71.174203} {\bibfield  {journal} {\bibinfo
  {journal} {Physical Review B}\ }\textbf {\bibinfo {volume} {71}},\ \bibinfo
  {pages} {174203} (\bibinfo {year} {2005})}\BibitemShut {NoStop}%
\bibitem [{\citenamefont {Celardo}\ \emph {et~al.}(2016)\citenamefont
  {Celardo}, \citenamefont {Kaiser},\ and\ \citenamefont
  {Borgonovi}}]{Celardo2016}%
  \BibitemOpen
  \bibfield  {author} {\bibinfo {author} {\bibfnamefont {G.~L.}\ \bibnamefont
  {Celardo}}, \bibinfo {author} {\bibfnamefont {R.}~\bibnamefont {Kaiser}},\
  and\ \bibinfo {author} {\bibfnamefont {F.}~\bibnamefont {Borgonovi}},\
  }\bibfield  {title} {\bibinfo {title} {Shielding and localization in the
  presence of long-range hopping},\ }\href
  {https://doi.org/10.1103/PhysRevB.94.144206} {\bibfield  {journal} {\bibinfo
  {journal} {Physical Review B}\ }\textbf {\bibinfo {volume} {94}},\ \bibinfo
  {pages} {144206} (\bibinfo {year} {2016})}\BibitemShut {NoStop}%
\bibitem [{\citenamefont {{de Paz}}\ \emph {et~al.}(2013)\citenamefont {{de
  Paz}}, \citenamefont {Sharma}, \citenamefont {Chotia}, \citenamefont
  {Mar{\'e}chal}, \citenamefont {Huckans}, \citenamefont {Pedri}, \citenamefont
  {Santos}, \citenamefont {Gorceix}, \citenamefont {Vernac},\ and\
  \citenamefont {{Laburthe-Tolra}}}]{dePaz2013}%
  \BibitemOpen
  \bibfield  {author} {\bibinfo {author} {\bibfnamefont {A.}~\bibnamefont {{de
  Paz}}}, \bibinfo {author} {\bibfnamefont {A.}~\bibnamefont {Sharma}},
  \bibinfo {author} {\bibfnamefont {A.}~\bibnamefont {Chotia}}, \bibinfo
  {author} {\bibfnamefont {E.}~\bibnamefont {Mar{\'e}chal}}, \bibinfo {author}
  {\bibfnamefont {J.~H.}\ \bibnamefont {Huckans}}, \bibinfo {author}
  {\bibfnamefont {P.}~\bibnamefont {Pedri}}, \bibinfo {author} {\bibfnamefont
  {L.}~\bibnamefont {Santos}}, \bibinfo {author} {\bibfnamefont
  {O.}~\bibnamefont {Gorceix}}, \bibinfo {author} {\bibfnamefont
  {L.}~\bibnamefont {Vernac}},\ and\ \bibinfo {author} {\bibfnamefont
  {B.}~\bibnamefont {{Laburthe-Tolra}}},\ }\bibfield  {title} {\bibinfo {title}
  {Nonequilibrium {{Quantum Magnetism}} in a {{Dipolar Lattice Gas}}},\ }\href
  {https://doi.org/10.1103/PhysRevLett.111.185305} {\bibfield  {journal}
  {\bibinfo  {journal} {Physical Review Letters}\ }\textbf {\bibinfo {volume}
  {111}},\ \bibinfo {pages} {185305} (\bibinfo {year} {2013})}\BibitemShut
  {NoStop}%
\bibitem [{\citenamefont {Baier}\ \emph {et~al.}(2016)\citenamefont {Baier},
  \citenamefont {Mark}, \citenamefont {Petter}, \citenamefont {Aikawa},
  \citenamefont {Chomaz}, \citenamefont {Cai}, \citenamefont {Baranov},
  \citenamefont {Zoller},\ and\ \citenamefont {Ferlaino}}]{Baier2016}%
  \BibitemOpen
  \bibfield  {author} {\bibinfo {author} {\bibfnamefont {S.}~\bibnamefont
  {Baier}}, \bibinfo {author} {\bibfnamefont {M.~J.}\ \bibnamefont {Mark}},
  \bibinfo {author} {\bibfnamefont {D.}~\bibnamefont {Petter}}, \bibinfo
  {author} {\bibfnamefont {K.}~\bibnamefont {Aikawa}}, \bibinfo {author}
  {\bibfnamefont {L.}~\bibnamefont {Chomaz}}, \bibinfo {author} {\bibfnamefont
  {Z.}~\bibnamefont {Cai}}, \bibinfo {author} {\bibfnamefont {M.}~\bibnamefont
  {Baranov}}, \bibinfo {author} {\bibfnamefont {P.}~\bibnamefont {Zoller}},\
  and\ \bibinfo {author} {\bibfnamefont {F.}~\bibnamefont {Ferlaino}},\
  }\bibfield  {title} {\bibinfo {title} {Extended {{Bose}}-{{Hubbard}} models
  with ultracold magnetic atoms},\ }\href
  {https://doi.org/10.1126/science.aac9812} {\bibfield  {journal} {\bibinfo
  {journal} {Science}\ }\textbf {\bibinfo {volume} {352}},\ \bibinfo {pages}
  {201} (\bibinfo {year} {2016})}\BibitemShut {NoStop}%
\bibitem [{\citenamefont {Lepoutre}\ \emph {et~al.}(2019)\citenamefont
  {Lepoutre}, \citenamefont {Schachenmayer}, \citenamefont {Gabardos},
  \citenamefont {Zhu}, \citenamefont {Naylor}, \citenamefont {Mar{\'e}chal},
  \citenamefont {Gorceix}, \citenamefont {Rey}, \citenamefont {Vernac},\ and\
  \citenamefont {{Laburthe-Tolra}}}]{Lepoutre2019}%
  \BibitemOpen
  \bibfield  {author} {\bibinfo {author} {\bibfnamefont {S.}~\bibnamefont
  {Lepoutre}}, \bibinfo {author} {\bibfnamefont {J.}~\bibnamefont
  {Schachenmayer}}, \bibinfo {author} {\bibfnamefont {L.}~\bibnamefont
  {Gabardos}}, \bibinfo {author} {\bibfnamefont {B.}~\bibnamefont {Zhu}},
  \bibinfo {author} {\bibfnamefont {B.}~\bibnamefont {Naylor}}, \bibinfo
  {author} {\bibfnamefont {E.}~\bibnamefont {Mar{\'e}chal}}, \bibinfo {author}
  {\bibfnamefont {O.}~\bibnamefont {Gorceix}}, \bibinfo {author} {\bibfnamefont
  {A.~M.}\ \bibnamefont {Rey}}, \bibinfo {author} {\bibfnamefont
  {L.}~\bibnamefont {Vernac}},\ and\ \bibinfo {author} {\bibfnamefont
  {B.}~\bibnamefont {{Laburthe-Tolra}}},\ }\bibfield  {title} {\bibinfo {title}
  {Out-of-equilibrium quantum magnetism and thermalization in a spin-3
  many-body dipolar lattice system},\ }\href
  {https://doi.org/10.1038/s41467-019-09699-5} {\bibfield  {journal} {\bibinfo
  {journal} {Nature Communications}\ }\textbf {\bibinfo {volume} {10}},\
  \bibinfo {pages} {1714} (\bibinfo {year} {2019})}\BibitemShut {NoStop}%
\bibitem [{\citenamefont {Patscheider}\ \emph {et~al.}(2020)\citenamefont
  {Patscheider}, \citenamefont {Zhu}, \citenamefont {Chomaz}, \citenamefont
  {Petter}, \citenamefont {Baier}, \citenamefont {Rey}, \citenamefont
  {Ferlaino},\ and\ \citenamefont {Mark}}]{Patscheider2020}%
  \BibitemOpen
  \bibfield  {author} {\bibinfo {author} {\bibfnamefont {A.}~\bibnamefont
  {Patscheider}}, \bibinfo {author} {\bibfnamefont {B.}~\bibnamefont {Zhu}},
  \bibinfo {author} {\bibfnamefont {L.}~\bibnamefont {Chomaz}}, \bibinfo
  {author} {\bibfnamefont {D.}~\bibnamefont {Petter}}, \bibinfo {author}
  {\bibfnamefont {S.}~\bibnamefont {Baier}}, \bibinfo {author} {\bibfnamefont
  {A.-M.}\ \bibnamefont {Rey}}, \bibinfo {author} {\bibfnamefont
  {F.}~\bibnamefont {Ferlaino}},\ and\ \bibinfo {author} {\bibfnamefont
  {M.~J.}\ \bibnamefont {Mark}},\ }\bibfield  {title} {\bibinfo {title}
  {Controlling dipolar exchange interactions in a dense three-dimensional array
  of large-spin fermions},\ }\href
  {https://doi.org/10.1103/PhysRevResearch.2.023050} {\bibfield  {journal}
  {\bibinfo  {journal} {Physical Review Research}\ }\textbf {\bibinfo {volume}
  {2}},\ \bibinfo {pages} {023050} (\bibinfo {year} {2020})}\BibitemShut
  {NoStop}%
\bibitem [{\citenamefont {Barredo}\ \emph {et~al.}(2015)\citenamefont
  {Barredo}, \citenamefont {Labuhn}, \citenamefont {Ravets}, \citenamefont
  {Lahaye}, \citenamefont {Browaeys},\ and\ \citenamefont
  {Adams}}]{Barredo2015}%
  \BibitemOpen
  \bibfield  {author} {\bibinfo {author} {\bibfnamefont {D.}~\bibnamefont
  {Barredo}}, \bibinfo {author} {\bibfnamefont {H.}~\bibnamefont {Labuhn}},
  \bibinfo {author} {\bibfnamefont {S.}~\bibnamefont {Ravets}}, \bibinfo
  {author} {\bibfnamefont {T.}~\bibnamefont {Lahaye}}, \bibinfo {author}
  {\bibfnamefont {A.}~\bibnamefont {Browaeys}},\ and\ \bibinfo {author}
  {\bibfnamefont {C.~S.}\ \bibnamefont {Adams}},\ }\bibfield  {title} {\bibinfo
  {title} {Coherent {{Excitation Transfer}} in a {{Spin Chain}} of {{Three
  Rydberg Atoms}}},\ }\href {https://doi.org/10.1103/PhysRevLett.114.113002}
  {\bibfield  {journal} {\bibinfo  {journal} {Physical Review Letters}\
  }\textbf {\bibinfo {volume} {114}},\ \bibinfo {pages} {113002} (\bibinfo
  {year} {2015})}\BibitemShut {NoStop}%
\bibitem [{\citenamefont {Orioli}\ \emph {et~al.}(2018)\citenamefont {Orioli},
  \citenamefont {Signoles}, \citenamefont {Wildhagen}, \citenamefont
  {G{\"u}nter}, \citenamefont {Berges}, \citenamefont {Whitlock},\ and\
  \citenamefont {Weidem{\"u}ller}}]{Orioli2018}%
  \BibitemOpen
  \bibfield  {author} {\bibinfo {author} {\bibfnamefont {A.~P.}\ \bibnamefont
  {Orioli}}, \bibinfo {author} {\bibfnamefont {A.}~\bibnamefont {Signoles}},
  \bibinfo {author} {\bibfnamefont {H.}~\bibnamefont {Wildhagen}}, \bibinfo
  {author} {\bibfnamefont {G.}~\bibnamefont {G{\"u}nter}}, \bibinfo {author}
  {\bibfnamefont {J.}~\bibnamefont {Berges}}, \bibinfo {author} {\bibfnamefont
  {S.}~\bibnamefont {Whitlock}},\ and\ \bibinfo {author} {\bibfnamefont
  {M.}~\bibnamefont {Weidem{\"u}ller}},\ }\bibfield  {title} {\bibinfo {title}
  {Relaxation of an {{Isolated Dipolar}}-{{Interacting Rydberg Quantum Spin
  System}}},\ }\href {https://doi.org/10.1103/PhysRevLett.120.063601}
  {\bibfield  {journal} {\bibinfo  {journal} {Physical Review Letters}\
  }\textbf {\bibinfo {volume} {120}},\ \bibinfo {pages} {063601} (\bibinfo
  {year} {2018})}\BibitemShut {NoStop}%
\bibitem [{\citenamefont {de~L{\'e}s{\'e}leuc}\ \emph
  {et~al.}(2019)\citenamefont {de~L{\'e}s{\'e}leuc}, \citenamefont {Lienhard},
  \citenamefont {Scholl}, \citenamefont {Barredo}, \citenamefont {Weber},
  \citenamefont {Lang}, \citenamefont {B{\"u}chler}, \citenamefont {Lahaye},\
  and\ \citenamefont {Browaeys}}]{Leseleuc2019}%
  \BibitemOpen
  \bibfield  {author} {\bibinfo {author} {\bibfnamefont {S.}~\bibnamefont
  {de~L{\'e}s{\'e}leuc}}, \bibinfo {author} {\bibfnamefont {V.}~\bibnamefont
  {Lienhard}}, \bibinfo {author} {\bibfnamefont {P.}~\bibnamefont {Scholl}},
  \bibinfo {author} {\bibfnamefont {D.}~\bibnamefont {Barredo}}, \bibinfo
  {author} {\bibfnamefont {S.}~\bibnamefont {Weber}}, \bibinfo {author}
  {\bibfnamefont {N.}~\bibnamefont {Lang}}, \bibinfo {author} {\bibfnamefont
  {H.~P.}\ \bibnamefont {B{\"u}chler}}, \bibinfo {author} {\bibfnamefont
  {T.}~\bibnamefont {Lahaye}},\ and\ \bibinfo {author} {\bibfnamefont
  {A.}~\bibnamefont {Browaeys}},\ }\bibfield  {title} {\bibinfo {title}
  {Observation of a symmetry-protected topological phase of interacting bosons
  with {{Rydberg}} atoms},\ }\href {https://doi.org/10.1126/science.aav9105}
  {\bibfield  {journal} {\bibinfo  {journal} {Science}\ }\textbf {\bibinfo
  {volume} {365}},\ \bibinfo {pages} {775} (\bibinfo {year}
  {2019})}\BibitemShut {NoStop}%
\bibitem [{\citenamefont {Richerme}\ \emph {et~al.}(2014)\citenamefont
  {Richerme}, \citenamefont {Gong}, \citenamefont {Lee}, \citenamefont {Senko},
  \citenamefont {Smith}, \citenamefont {{Foss-Feig}}, \citenamefont
  {Michalakis}, \citenamefont {Gorshkov},\ and\ \citenamefont
  {Monroe}}]{Richerme2014}%
  \BibitemOpen
  \bibfield  {author} {\bibinfo {author} {\bibfnamefont {P.}~\bibnamefont
  {Richerme}}, \bibinfo {author} {\bibfnamefont {Z.-X.}\ \bibnamefont {Gong}},
  \bibinfo {author} {\bibfnamefont {A.}~\bibnamefont {Lee}}, \bibinfo {author}
  {\bibfnamefont {C.}~\bibnamefont {Senko}}, \bibinfo {author} {\bibfnamefont
  {J.}~\bibnamefont {Smith}}, \bibinfo {author} {\bibfnamefont
  {M.}~\bibnamefont {{Foss-Feig}}}, \bibinfo {author} {\bibfnamefont
  {S.}~\bibnamefont {Michalakis}}, \bibinfo {author} {\bibfnamefont {A.~V.}\
  \bibnamefont {Gorshkov}},\ and\ \bibinfo {author} {\bibfnamefont
  {C.}~\bibnamefont {Monroe}},\ }\bibfield  {title} {\bibinfo {title}
  {Non-local propagation of correlations in quantum systems with long-range
  interactions},\ }\href {https://doi.org/10.1038/nature13450} {\bibfield
  {journal} {\bibinfo  {journal} {Nature}\ }\textbf {\bibinfo {volume} {511}},\
  \bibinfo {pages} {198} (\bibinfo {year} {2014})}\BibitemShut {NoStop}%
\bibitem [{\citenamefont {Jurcevic}\ \emph {et~al.}(2014)\citenamefont
  {Jurcevic}, \citenamefont {Lanyon}, \citenamefont {Hauke}, \citenamefont
  {Hempel}, \citenamefont {Zoller}, \citenamefont {Blatt},\ and\ \citenamefont
  {Roos}}]{Jurcevic2014}%
  \BibitemOpen
  \bibfield  {author} {\bibinfo {author} {\bibfnamefont {P.}~\bibnamefont
  {Jurcevic}}, \bibinfo {author} {\bibfnamefont {B.~P.}\ \bibnamefont
  {Lanyon}}, \bibinfo {author} {\bibfnamefont {P.}~\bibnamefont {Hauke}},
  \bibinfo {author} {\bibfnamefont {C.}~\bibnamefont {Hempel}}, \bibinfo
  {author} {\bibfnamefont {P.}~\bibnamefont {Zoller}}, \bibinfo {author}
  {\bibfnamefont {R.}~\bibnamefont {Blatt}},\ and\ \bibinfo {author}
  {\bibfnamefont {C.~F.}\ \bibnamefont {Roos}},\ }\bibfield  {title} {\bibinfo
  {title} {Quasiparticle engineering and entanglement propagation in a quantum
  many-body system},\ }\href {https://doi.org/10.1038/nature13461} {\bibfield
  {journal} {\bibinfo  {journal} {Nature}\ }\textbf {\bibinfo {volume} {511}},\
  \bibinfo {pages} {202} (\bibinfo {year} {2014})}\BibitemShut {NoStop}%
\bibitem [{\citenamefont {Yan}\ \emph {et~al.}(2013)\citenamefont {Yan},
  \citenamefont {Moses}, \citenamefont {Gadway}, \citenamefont {Covey},
  \citenamefont {Hazzard}, \citenamefont {Rey}, \citenamefont {Jin},\ and\
  \citenamefont {Ye}}]{Yan2013}%
  \BibitemOpen
  \bibfield  {author} {\bibinfo {author} {\bibfnamefont {B.}~\bibnamefont
  {Yan}}, \bibinfo {author} {\bibfnamefont {S.~A.}\ \bibnamefont {Moses}},
  \bibinfo {author} {\bibfnamefont {B.}~\bibnamefont {Gadway}}, \bibinfo
  {author} {\bibfnamefont {J.~P.}\ \bibnamefont {Covey}}, \bibinfo {author}
  {\bibfnamefont {K.~R.~A.}\ \bibnamefont {Hazzard}}, \bibinfo {author}
  {\bibfnamefont {A.~M.}\ \bibnamefont {Rey}}, \bibinfo {author} {\bibfnamefont
  {D.~S.}\ \bibnamefont {Jin}},\ and\ \bibinfo {author} {\bibfnamefont
  {J.}~\bibnamefont {Ye}},\ }\bibfield  {title} {\bibinfo {title} {Observation
  of dipolar spin-exchange interactions with lattice-confined polar
  molecules},\ }\href {https://doi.org/10.1038/nature12483} {\bibfield
  {journal} {\bibinfo  {journal} {Nature}\ }\textbf {\bibinfo {volume} {501}},\
  \bibinfo {pages} {521} (\bibinfo {year} {2013})}\BibitemShut {NoStop}%
\bibitem [{\citenamefont {{\'A}lvarez}\ \emph {et~al.}(2015)\citenamefont
  {{\'A}lvarez}, \citenamefont {Suter},\ and\ \citenamefont
  {Kaiser}}]{Alvarez2015}%
  \BibitemOpen
  \bibfield  {author} {\bibinfo {author} {\bibfnamefont {G.~A.}\ \bibnamefont
  {{\'A}lvarez}}, \bibinfo {author} {\bibfnamefont {D.}~\bibnamefont {Suter}},\
  and\ \bibinfo {author} {\bibfnamefont {R.}~\bibnamefont {Kaiser}},\
  }\bibfield  {title} {\bibinfo {title} {Localization-delocalization transition
  in the dynamics of dipolar-coupled nuclear spins},\ }\href
  {https://doi.org/10.1126/science.1261160} {\bibfield  {journal} {\bibinfo
  {journal} {Science}\ }\textbf {\bibinfo {volume} {349}},\ \bibinfo {pages}
  {846} (\bibinfo {year} {2015})}\BibitemShut {NoStop}%
\bibitem [{\citenamefont {Waldherr}\ \emph {et~al.}(2014)\citenamefont
  {Waldherr}, \citenamefont {Wang}, \citenamefont {Zaiser}, \citenamefont
  {Jamali}, \citenamefont {{Schulte-Herbr{\"u}ggen}}, \citenamefont {Abe},
  \citenamefont {Ohshima}, \citenamefont {Isoya}, \citenamefont {Du},
  \citenamefont {Neumann},\ and\ \citenamefont {Wrachtrup}}]{Waldherr2014}%
  \BibitemOpen
  \bibfield  {author} {\bibinfo {author} {\bibfnamefont {G.}~\bibnamefont
  {Waldherr}}, \bibinfo {author} {\bibfnamefont {Y.}~\bibnamefont {Wang}},
  \bibinfo {author} {\bibfnamefont {S.}~\bibnamefont {Zaiser}}, \bibinfo
  {author} {\bibfnamefont {M.}~\bibnamefont {Jamali}}, \bibinfo {author}
  {\bibfnamefont {T.}~\bibnamefont {{Schulte-Herbr{\"u}ggen}}}, \bibinfo
  {author} {\bibfnamefont {H.}~\bibnamefont {Abe}}, \bibinfo {author}
  {\bibfnamefont {T.}~\bibnamefont {Ohshima}}, \bibinfo {author} {\bibfnamefont
  {J.}~\bibnamefont {Isoya}}, \bibinfo {author} {\bibfnamefont {J.~F.}\
  \bibnamefont {Du}}, \bibinfo {author} {\bibfnamefont {P.}~\bibnamefont
  {Neumann}},\ and\ \bibinfo {author} {\bibfnamefont {J.}~\bibnamefont
  {Wrachtrup}},\ }\bibfield  {title} {\bibinfo {title} {Quantum error
  correction in a solid-state hybrid spin register},\ }\href
  {https://doi.org/10.1038/nature12919} {\bibfield  {journal} {\bibinfo
  {journal} {Nature}\ }\textbf {\bibinfo {volume} {506}},\ \bibinfo {pages}
  {204} (\bibinfo {year} {2014})}\BibitemShut {NoStop}%
\bibitem [{\citenamefont {Hung}\ \emph {et~al.}(2016)\citenamefont {Hung},
  \citenamefont {{Gonz{\'a}lez-Tudela}}, \citenamefont {Cirac},\ and\
  \citenamefont {Kimble}}]{Hung2016}%
  \BibitemOpen
  \bibfield  {author} {\bibinfo {author} {\bibfnamefont {C.-L.}\ \bibnamefont
  {Hung}}, \bibinfo {author} {\bibfnamefont {A.}~\bibnamefont
  {{Gonz{\'a}lez-Tudela}}}, \bibinfo {author} {\bibfnamefont {J.~I.}\
  \bibnamefont {Cirac}},\ and\ \bibinfo {author} {\bibfnamefont {H.~J.}\
  \bibnamefont {Kimble}},\ }\bibfield  {title} {\bibinfo {title} {Quantum spin
  dynamics with pairwise-tunable, long-range interactions},\ }\href
  {https://doi.org/10.1073/pnas.1603777113} {\bibfield  {journal} {\bibinfo
  {journal} {Proceedings of the National Academy of Sciences}\ }\textbf
  {\bibinfo {volume} {113}},\ \bibinfo {pages} {E4946} (\bibinfo {year}
  {2016})}\BibitemShut {NoStop}%
\bibitem [{\citenamefont {Feigelman}\ \emph {et~al.}(1993)\citenamefont
  {Feigelman}, \citenamefont {Geshkenbein}, \citenamefont {Ioffe},\ and\
  \citenamefont {Larkin}}]{Feigelman1993}%
  \BibitemOpen
  \bibfield  {author} {\bibinfo {author} {\bibfnamefont {M.~V.}\ \bibnamefont
  {Feigelman}}, \bibinfo {author} {\bibfnamefont {V.~B.}\ \bibnamefont
  {Geshkenbein}}, \bibinfo {author} {\bibfnamefont {L.~B.}\ \bibnamefont
  {Ioffe}},\ and\ \bibinfo {author} {\bibfnamefont {A.~I.}\ \bibnamefont
  {Larkin}},\ }\bibfield  {title} {\bibinfo {title} {Two-dimensional {{Bose}}
  liquid with strong gauge-field interaction},\ }\href
  {https://doi.org/10.1103/PhysRevB.48.16641} {\bibfield  {journal} {\bibinfo
  {journal} {Physical Review B}\ }\textbf {\bibinfo {volume} {48}},\ \bibinfo
  {pages} {16641} (\bibinfo {year} {1993})}\BibitemShut {NoStop}%
\bibitem [{\citenamefont {Phillips}\ and\ \citenamefont
  {Dalidovich}(2003)}]{Phillips2003}%
  \BibitemOpen
  \bibfield  {author} {\bibinfo {author} {\bibfnamefont {P.}~\bibnamefont
  {Phillips}}\ and\ \bibinfo {author} {\bibfnamefont {D.}~\bibnamefont
  {Dalidovich}},\ }\bibfield  {title} {\bibinfo {title} {The {{Elusive Bose
  Metal}}},\ }\href {https://doi.org/10.1126/science.1088253} {\bibfield
  {journal} {\bibinfo  {journal} {Science}\ }\textbf {\bibinfo {volume}
  {302}},\ \bibinfo {pages} {243} (\bibinfo {year} {2003})}\BibitemShut
  {NoStop}%
\bibitem [{\citenamefont {Motrunich}\ and\ \citenamefont
  {Fisher}(2007)}]{Motrunich2007}%
  \BibitemOpen
  \bibfield  {author} {\bibinfo {author} {\bibfnamefont {O.~I.}\ \bibnamefont
  {Motrunich}}\ and\ \bibinfo {author} {\bibfnamefont {M.~P.~A.}\ \bibnamefont
  {Fisher}},\ }\bibfield  {title} {\bibinfo {title} {\$d\$-wave correlated
  critical {{Bose}} liquids in two dimensions},\ }\href
  {https://doi.org/10.1103/PhysRevB.75.235116} {\bibfield  {journal} {\bibinfo
  {journal} {Physical Review B}\ }\textbf {\bibinfo {volume} {75}},\ \bibinfo
  {pages} {235116} (\bibinfo {year} {2007})}\BibitemShut {NoStop}%
\bibitem [{\citenamefont {Yang}\ \emph {et~al.}(2019)\citenamefont {Yang},
  \citenamefont {Liu}, \citenamefont {Wang}, \citenamefont {Feng},
  \citenamefont {He}, \citenamefont {Sun}, \citenamefont {Tang}, \citenamefont
  {Wu}, \citenamefont {Xiong}, \citenamefont {Zhang}, \citenamefont {Lin},
  \citenamefont {Yao}, \citenamefont {Liu}, \citenamefont {Fernandes},
  \citenamefont {Xu}, \citenamefont {Valles}, \citenamefont {Wang},\ and\
  \citenamefont {Li}}]{Yang2019}%
  \BibitemOpen
  \bibfield  {author} {\bibinfo {author} {\bibfnamefont {C.}~\bibnamefont
  {Yang}}, \bibinfo {author} {\bibfnamefont {Y.}~\bibnamefont {Liu}}, \bibinfo
  {author} {\bibfnamefont {Y.}~\bibnamefont {Wang}}, \bibinfo {author}
  {\bibfnamefont {L.}~\bibnamefont {Feng}}, \bibinfo {author} {\bibfnamefont
  {Q.}~\bibnamefont {He}}, \bibinfo {author} {\bibfnamefont {J.}~\bibnamefont
  {Sun}}, \bibinfo {author} {\bibfnamefont {Y.}~\bibnamefont {Tang}}, \bibinfo
  {author} {\bibfnamefont {C.}~\bibnamefont {Wu}}, \bibinfo {author}
  {\bibfnamefont {J.}~\bibnamefont {Xiong}}, \bibinfo {author} {\bibfnamefont
  {W.}~\bibnamefont {Zhang}}, \bibinfo {author} {\bibfnamefont
  {X.}~\bibnamefont {Lin}}, \bibinfo {author} {\bibfnamefont {H.}~\bibnamefont
  {Yao}}, \bibinfo {author} {\bibfnamefont {H.}~\bibnamefont {Liu}}, \bibinfo
  {author} {\bibfnamefont {G.}~\bibnamefont {Fernandes}}, \bibinfo {author}
  {\bibfnamefont {J.}~\bibnamefont {Xu}}, \bibinfo {author} {\bibfnamefont
  {J.~M.}\ \bibnamefont {Valles}}, \bibinfo {author} {\bibfnamefont
  {J.}~\bibnamefont {Wang}},\ and\ \bibinfo {author} {\bibfnamefont
  {Y.}~\bibnamefont {Li}},\ }\bibfield  {title} {\bibinfo {title} {Intermediate
  bosonic metallic state in the superconductor-insulator transition},\ }\href
  {https://doi.org/10.1126/science.aax5798} {\bibfield  {journal} {\bibinfo
  {journal} {Science}\ }\textbf {\bibinfo {volume} {366}},\ \bibinfo {pages}
  {1505} (\bibinfo {year} {2019})}\BibitemShut {NoStop}%
\bibitem [{\citenamefont {Jiang}\ \emph {et~al.}(2013)\citenamefont {Jiang},
  \citenamefont {Block}, \citenamefont {Mishmash}, \citenamefont {Garrison},
  \citenamefont {Sheng}, \citenamefont {Motrunich},\ and\ \citenamefont
  {Fisher}}]{Jiang2013}%
  \BibitemOpen
  \bibfield  {author} {\bibinfo {author} {\bibfnamefont {H.-C.}\ \bibnamefont
  {Jiang}}, \bibinfo {author} {\bibfnamefont {M.~S.}\ \bibnamefont {Block}},
  \bibinfo {author} {\bibfnamefont {R.~V.}\ \bibnamefont {Mishmash}}, \bibinfo
  {author} {\bibfnamefont {J.~R.}\ \bibnamefont {Garrison}}, \bibinfo {author}
  {\bibfnamefont {D.~N.}\ \bibnamefont {Sheng}}, \bibinfo {author}
  {\bibfnamefont {O.~I.}\ \bibnamefont {Motrunich}},\ and\ \bibinfo {author}
  {\bibfnamefont {M.~P.~A.}\ \bibnamefont {Fisher}},\ }\bibfield  {title}
  {\bibinfo {title} {Non-{{Fermi}}-liquid d -wave metal phase of strongly
  interacting electrons},\ }\href {https://doi.org/10.1038/nature11732}
  {\bibfield  {journal} {\bibinfo  {journal} {Nature}\ }\textbf {\bibinfo
  {volume} {493}},\ \bibinfo {pages} {39} (\bibinfo {year} {2013})}\BibitemShut
  {NoStop}%
\bibitem [{\citenamefont {Block}\ \emph {et~al.}(2011)\citenamefont {Block},
  \citenamefont {Sheng}, \citenamefont {Motrunich},\ and\ \citenamefont
  {Fisher}}]{Block2011}%
  \BibitemOpen
  \bibfield  {author} {\bibinfo {author} {\bibfnamefont {M.~S.}\ \bibnamefont
  {Block}}, \bibinfo {author} {\bibfnamefont {D.~N.}\ \bibnamefont {Sheng}},
  \bibinfo {author} {\bibfnamefont {O.~I.}\ \bibnamefont {Motrunich}},\ and\
  \bibinfo {author} {\bibfnamefont {M.~P.~A.}\ \bibnamefont {Fisher}},\
  }\bibfield  {title} {\bibinfo {title} {Spin {{Bose}}-{{Metal}} and {{Valence
  Bond Solid Phases}} in a {{Spin}}-\$1/2\$ {{Model}} with {{Ring Exchanges}}
  on a {{Four}}-{{Leg Triangular Ladder}}},\ }\href
  {https://doi.org/10.1103/PhysRevLett.106.157202} {\bibfield  {journal}
  {\bibinfo  {journal} {Physical Review Letters}\ }\textbf {\bibinfo {volume}
  {106}},\ \bibinfo {pages} {157202} (\bibinfo {year} {2011})}\BibitemShut
  {NoStop}%
\bibitem [{\citenamefont {Zeiher}\ \emph {et~al.}(2017)\citenamefont {Zeiher},
  \citenamefont {Choi}, \citenamefont {{Rubio-Abadal}}, \citenamefont {Pohl},
  \citenamefont {{van Bijnen}}, \citenamefont {Bloch},\ and\ \citenamefont
  {Gross}}]{Zeiher2017}%
  \BibitemOpen
  \bibfield  {author} {\bibinfo {author} {\bibfnamefont {J.}~\bibnamefont
  {Zeiher}}, \bibinfo {author} {\bibfnamefont {J.-y.}\ \bibnamefont {Choi}},
  \bibinfo {author} {\bibfnamefont {A.}~\bibnamefont {{Rubio-Abadal}}},
  \bibinfo {author} {\bibfnamefont {T.}~\bibnamefont {Pohl}}, \bibinfo {author}
  {\bibfnamefont {R.}~\bibnamefont {{van Bijnen}}}, \bibinfo {author}
  {\bibfnamefont {I.}~\bibnamefont {Bloch}},\ and\ \bibinfo {author}
  {\bibfnamefont {C.}~\bibnamefont {Gross}},\ }\bibfield  {title} {\bibinfo
  {title} {Coherent {{Many}}-{{Body Spin Dynamics}} in a {{Long}}-{{Range
  Interacting Ising Chain}}},\ }\href
  {https://doi.org/10.1103/PhysRevX.7.041063} {\bibfield  {journal} {\bibinfo
  {journal} {Physical Review X}\ }\textbf {\bibinfo {volume} {7}},\ \bibinfo
  {pages} {041063} (\bibinfo {year} {2017})}\BibitemShut {NoStop}%
\bibitem [{\citenamefont {Burin}(2015{\natexlab{a}})}]{Burin2015}%
  \BibitemOpen
  \bibfield  {author} {\bibinfo {author} {\bibfnamefont {A.~L.}\ \bibnamefont
  {Burin}},\ }\bibfield  {title} {\bibinfo {title} {Localization in a random
  {{XY}} model with long-range interactions: {{Intermediate}} case between
  single-particle and many-body problems},\ }\href
  {https://doi.org/10.1103/PhysRevB.92.104428} {\bibfield  {journal} {\bibinfo
  {journal} {Physical Review B}\ }\textbf {\bibinfo {volume} {92}},\ \bibinfo
  {pages} {104428} (\bibinfo {year} {2015}{\natexlab{a}})}\BibitemShut
  {NoStop}%
\bibitem [{\citenamefont {Burin}(2015{\natexlab{b}})}]{Burin2015a}%
  \BibitemOpen
  \bibfield  {author} {\bibinfo {author} {\bibfnamefont {A.~L.}\ \bibnamefont
  {Burin}},\ }\bibfield  {title} {\bibinfo {title} {Many-body delocalization in
  a strongly disordered system with long-range interactions: {{Finite}}-size
  scaling},\ }\href {https://doi.org/10.1103/PhysRevB.91.094202} {\bibfield
  {journal} {\bibinfo  {journal} {Physical Review B}\ }\textbf {\bibinfo
  {volume} {91}},\ \bibinfo {pages} {094202} (\bibinfo {year}
  {2015}{\natexlab{b}})}\BibitemShut {NoStop}%
\bibitem [{\citenamefont {{Safavi-Naini}}\ \emph {et~al.}(2019)\citenamefont
  {{Safavi-Naini}}, \citenamefont {Wall}, \citenamefont {Acevedo},
  \citenamefont {Rey},\ and\ \citenamefont {Nandkishore}}]{Safavi-Naini2019}%
  \BibitemOpen
  \bibfield  {author} {\bibinfo {author} {\bibfnamefont {A.}~\bibnamefont
  {{Safavi-Naini}}}, \bibinfo {author} {\bibfnamefont {M.~L.}\ \bibnamefont
  {Wall}}, \bibinfo {author} {\bibfnamefont {O.~L.}\ \bibnamefont {Acevedo}},
  \bibinfo {author} {\bibfnamefont {A.~M.}\ \bibnamefont {Rey}},\ and\ \bibinfo
  {author} {\bibfnamefont {R.~M.}\ \bibnamefont {Nandkishore}},\ }\bibfield
  {title} {\bibinfo {title} {Quantum dynamics of disordered spin chains with
  power-law interactions},\ }\href {https://doi.org/10.1103/PhysRevA.99.033610}
  {\bibfield  {journal} {\bibinfo  {journal} {Physical Review A}\ }\textbf
  {\bibinfo {volume} {99}},\ \bibinfo {pages} {033610} (\bibinfo {year}
  {2019})}\BibitemShut {NoStop}%
\bibitem [{\citenamefont {Deng}\ \emph {et~al.}(2020)\citenamefont {Deng},
  \citenamefont {Masella}, \citenamefont {Pupillo},\ and\ \citenamefont
  {Santos}}]{Deng2020}%
  \BibitemOpen
  \bibfield  {author} {\bibinfo {author} {\bibfnamefont {X.}~\bibnamefont
  {Deng}}, \bibinfo {author} {\bibfnamefont {G.}~\bibnamefont {Masella}},
  \bibinfo {author} {\bibfnamefont {G.}~\bibnamefont {Pupillo}},\ and\ \bibinfo
  {author} {\bibfnamefont {L.}~\bibnamefont {Santos}},\ }\bibfield  {title}
  {\bibinfo {title} {Universal {{Algebraic Growth}} of {{Entanglement Entropy}}
  in {{Many}}-{{Body Localized Systems}} with {{Power}}-{{Law Interactions}}},\
  }\href {https://doi.org/10.1103/PhysRevLett.125.010401} {\bibfield  {journal}
  {\bibinfo  {journal} {Physical Review Letters}\ }\textbf {\bibinfo {volume}
  {125}},\ \bibinfo {pages} {010401} (\bibinfo {year} {2020})}\BibitemShut
  {NoStop}%
\bibitem [{\citenamefont {Kagan}\ and\ \citenamefont
  {Maksimov}(1983)}]{Kagan1983}%
  \BibitemOpen
  \bibfield  {author} {\bibinfo {author} {\bibfnamefont {Y.}~\bibnamefont
  {Kagan}}\ and\ \bibinfo {author} {\bibfnamefont {L.~A.}\ \bibnamefont
  {Maksimov}},\ }\bibfield  {title} {\bibinfo {title} {Quantum diffusion of
  atoms in a crystal localization and phonon-stimulated delocalization},\
  }\href {https://doi.org/10.1016/0375-9601(83)90616-3} {\bibfield  {journal}
  {\bibinfo  {journal} {Physics Letters A}\ }\textbf {\bibinfo {volume} {95}},\
  \bibinfo {pages} {242} (\bibinfo {year} {1983})}\BibitemShut {NoStop}%
\bibitem [{\citenamefont {Prokof'ev}\ \emph {et~al.}(1998)\citenamefont
  {Prokof'ev}, \citenamefont {Svistunov},\ and\ \citenamefont
  {Tupitsyn}}]{Prokofev1998}%
  \BibitemOpen
  \bibfield  {author} {\bibinfo {author} {\bibfnamefont {N.~V.}\ \bibnamefont
  {Prokof'ev}}, \bibinfo {author} {\bibfnamefont {B.~V.}\ \bibnamefont
  {Svistunov}},\ and\ \bibinfo {author} {\bibfnamefont {I.~S.}\ \bibnamefont
  {Tupitsyn}},\ }\bibfield  {title} {\bibinfo {title} {Exact, complete, and
  universal continuous-time worldline {{Monte Carlo}} approach to the
  statistics of discrete quantum systems},\ }\href
  {https://doi.org/10.1134/1.558661} {\bibfield  {journal} {\bibinfo  {journal}
  {Journal of Experimental and Theoretical Physics}\ }\textbf {\bibinfo
  {volume} {87}},\ \bibinfo {pages} {310} (\bibinfo {year} {1998})}\BibitemShut
  {NoStop}%
\bibitem [{\citenamefont {Pollock}\ and\ \citenamefont
  {Ceperley}(1987)}]{Pollock1987}%
  \BibitemOpen
  \bibfield  {author} {\bibinfo {author} {\bibfnamefont {E.~L.}\ \bibnamefont
  {Pollock}}\ and\ \bibinfo {author} {\bibfnamefont {D.~M.}\ \bibnamefont
  {Ceperley}},\ }\bibfield  {title} {\bibinfo {title} {Path-integral
  computation of superfluid densities},\ }\href
  {https://doi.org/10.1103/PhysRevB.36.8343} {\bibfield  {journal} {\bibinfo
  {journal} {Physical Review B}\ }\textbf {\bibinfo {volume} {36}},\ \bibinfo
  {pages} {8343} (\bibinfo {year} {1987})}\BibitemShut {NoStop}%
\bibitem [{Note1()}]{Note1}%
  \BibitemOpen
  \bibinfo {note} {See Supplementary material}\BibitemShut {NoStop}%
\bibitem [{\citenamefont {Svistunov}\ \emph {et~al.}(2015)\citenamefont
  {Svistunov}, \citenamefont {Babaev},\ and\ \citenamefont
  {Prokof'ev}}]{Svistunov2015}%
  \BibitemOpen
  \bibfield  {author} {\bibinfo {author} {\bibfnamefont {B.~V.}\ \bibnamefont
  {Svistunov}}, \bibinfo {author} {\bibfnamefont {E.}~\bibnamefont {Babaev}},\
  and\ \bibinfo {author} {\bibfnamefont {N.~V.}\ \bibnamefont {Prokof'ev}},\
  }\href@noop {} {\emph {\bibinfo {title} {Superfluid {{States}} of
  {{Matter}}}}},\ \bibinfo {edition} {1st}\ ed.\ (\bibinfo  {publisher} {{CRC
  Press}},\ \bibinfo {year} {2015})\BibitemShut {NoStop}%
\bibitem [{\citenamefont {Levy}\ \emph {et~al.}(2017)\citenamefont {Levy},
  \citenamefont {LeBlanc},\ and\ \citenamefont {Gull}}]{Levy2017}%
  \BibitemOpen
  \bibfield  {author} {\bibinfo {author} {\bibfnamefont {R.}~\bibnamefont
  {Levy}}, \bibinfo {author} {\bibfnamefont {J.~P.~F.}\ \bibnamefont
  {LeBlanc}},\ and\ \bibinfo {author} {\bibfnamefont {E.}~\bibnamefont
  {Gull}},\ }\bibfield  {title} {\bibinfo {title} {Implementation of the
  maximum entropy method for analytic continuation},\ }\href
  {https://doi.org/10.1016/j.cpc.2017.01.018} {\bibfield  {journal} {\bibinfo
  {journal} {Computer Physics Communications}\ }\textbf {\bibinfo {volume}
  {215}},\ \bibinfo {pages} {149} (\bibinfo {year} {2017})}\BibitemShut
  {NoStop}%
\bibitem [{\citenamefont {Prokof'ev}\ and\ \citenamefont
  {Svistunov}(2013)}]{Prokofev2013a}%
  \BibitemOpen
  \bibfield  {author} {\bibinfo {author} {\bibfnamefont {N.~V.}\ \bibnamefont
  {Prokof'ev}}\ and\ \bibinfo {author} {\bibfnamefont {B.~V.}\ \bibnamefont
  {Svistunov}},\ }\bibfield  {title} {\bibinfo {title} {Spectral analysis by
  the method of consistent constraints},\ }\href
  {https://doi.org/10.1134/S002136401311009X} {\bibfield  {journal} {\bibinfo
  {journal} {JETP Letters}\ }\textbf {\bibinfo {volume} {97}},\ \bibinfo
  {pages} {649} (\bibinfo {year} {2013})}\BibitemShut {NoStop}%
\bibitem [{\citenamefont {Goulko}\ \emph {et~al.}(2017)\citenamefont {Goulko},
  \citenamefont {Mishchenko}, \citenamefont {Pollet}, \citenamefont
  {Prokof'ev},\ and\ \citenamefont {Svistunov}}]{Goulko2017}%
  \BibitemOpen
  \bibfield  {author} {\bibinfo {author} {\bibfnamefont {O.}~\bibnamefont
  {Goulko}}, \bibinfo {author} {\bibfnamefont {A.~S.}\ \bibnamefont
  {Mishchenko}}, \bibinfo {author} {\bibfnamefont {L.}~\bibnamefont {Pollet}},
  \bibinfo {author} {\bibfnamefont {N.}~\bibnamefont {Prokof'ev}},\ and\
  \bibinfo {author} {\bibfnamefont {B.}~\bibnamefont {Svistunov}},\ }\bibfield
  {title} {\bibinfo {title} {Numerical analytic continuation: {{Answers}} to
  well-posed questions},\ }\href {https://doi.org/10.1103/PhysRevB.95.014102}
  {\bibfield  {journal} {\bibinfo  {journal} {Physical Review B}\ }\textbf
  {\bibinfo {volume} {95}},\ \bibinfo {pages} {014102} (\bibinfo {year}
  {2017})}\BibitemShut {NoStop}%
\bibitem [{\citenamefont {Jarrell}\ and\ \citenamefont
  {Gubernatis}(1996)}]{Jarrell1996}%
  \BibitemOpen
  \bibfield  {author} {\bibinfo {author} {\bibfnamefont {M.}~\bibnamefont
  {Jarrell}}\ and\ \bibinfo {author} {\bibfnamefont {J.~E.}\ \bibnamefont
  {Gubernatis}},\ }\bibfield  {title} {\bibinfo {title} {Bayesian inference and
  the analytic continuation of imaginary-time quantum {{Monte Carlo}} data},\
  }\href {https://doi.org/10.1016/0370-1573(95)00074-7} {\bibfield  {journal}
  {\bibinfo  {journal} {Physics Reports}\ }\textbf {\bibinfo {volume} {269}},\
  \bibinfo {pages} {133} (\bibinfo {year} {1996})}\BibitemShut {NoStop}%
\end{thebibliography}%

\clearpage

\section*{Methods}
We perform quantum Monte Carlo simulations of Hamiltonian \cref{eq:hamiltonian}
in the path-integral representation in the grand-canonical ensemble using the
worm algorithm \cite{Prokofev1998} for system sizes as large as $L=256$ and
temperatures as low as $T/t = 1/256$. At half-filling, we shift disorder
realizations to ensue that $\ev{W_i} = \mu = 0$ for each realization, with
$\mu$ the chemical potential.  The resulting density is then $\ev{\rho} =
\frac{1}{2}$ when averaged over the disorder realizations with tiny, i.e.
$2.8\%$ for $L=256$ and $W=6.0$, sample-to-sample fluctuations.

In the presence of a constant vector potential \cref{eq:hamiltonian} is
modified by phase factors in the hopping elements of the form $t_{ij}
\rightarrow \mathrm{e}^{\imath \phi r_{ij}}$.  An expansion of the phase factor
up to the second order in $\phi$ leads to the current operator for the studied
Hamiltonian
\begin{equation}
    j = \imath t \sum_{i<j} \frac{r_{ij}}{\abs{r_{ij}}^3} \bqty{\cre{i}\ann{j}
    - \cre{j}\ann{i}}
\end{equation}
along with the additional operator $ \mathcal{T}$ that is required for proper
definition of the current-current correlation function (see below)
\begin{equation}
    \mathcal{T} = - t \sum_{i<j} \frac{r_{ij}^2}{\abs{r_{ij}}^3}
    \bqty{\cre{i}\ann{j} + \cre{j}\ann{i}}.
\end{equation}
The superfluid stiffness is, as usual, defined as the response of the free
energy $F$ to a weak externally applied phase $\phi$
\begin{equation}
    \Upsilon_s = L \left.\pdv[2]{F(\phi)}{\phi}\right|_{\phi=0},
\end{equation}
which in quantum Monte Carlo calculations is directly computed
as~\cite{Pollock1987,Svistunov2015}
\begin{equation}
    \Upsilon_s = L \ev{\mathcal{W}^2},
\end{equation}
with $\mathcal{W}$  the path winding number. In the case of hopping connecting
distant sites, as in \cref{eq:hamiltonian}, $\mathcal{W}$ can be written as
$\mathcal{W} = N_{\rightarrow} - N_{\leftarrow} = \sum_k r_k$ with
$N_{\leftrightarrows}$ the number of particle trajectories crossing the
hypothetical boundary of the system in a given direction, and with the sum
going over all the hopping elements in a single worldline configuration of the
entire system (here, $r_k$ represents the displacement between the sites
connected by the $k$-th hopping event).

\paragraph{Current-current correlation functions}
In the regime of weak field $\phi$ (linear response) it is sufficient to look
at the current-current correlation function
\begin{equation}
    \chi(\imath\omega_n) = \ev{j(\tau) j(0)}_{\imath\omega_n}
\end{equation}
at Matsubara frequencies $\omega_n = 2\pi T n$ ($n > 0$).  We compute it
numerically and perform an analytic continuation procedure to obtain the
conductivity $\sigma(\omega)$.
Here, the subscript $\imath \omega_n$ denotes that the Fourier transform is
taken of the corresponding correlation function $\ev{j(\tau) j(0)}$ in
imaginary time.
Path integral representation of quantum statistics for the Hamiltonian
\cref{eq:hamiltonian} allows one to sample Fourier components of this
correlation function directly, and collect statistics for different Matsubara
frequencies by using the  estimator $\abs{\sum_k \imath r_k
\mathrm{e}^{\imath\omega_n \tau_k}}^2$, where again the sum goes over all
hopping transitions on the systems worldlines.
For zero-frequency \(\omega_n=0\),  this estimator is equivalent to measuring
the winding number squared \(\mathcal{W}^2\) while for large Matsubara
frequencies it approaches the constant value corresponding to the estimator for
\(\mathcal{T}\).  After computing statistical averages, we subtract
\(\ev{\mathcal{T}}\) from the data to obtain the current-current correlation
function.
To suppress finite size effects associated with rare configurations with finite
winding numbers, we restrict the sampling of the correlation function
$\chi(\imath\omega_n)$ to configurations $\mathcal{W} = 0$.

\begin{figure}
    \centering
    \includegraphics[width=\columnwidth]{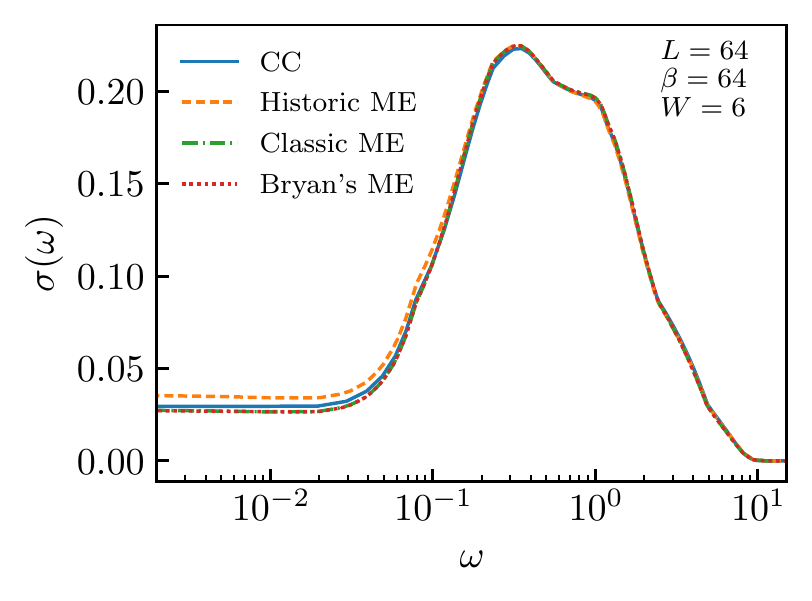}
    \caption{
        \label{fig:maxent_avg}
        Averaged optical conductivity $\sigma$ as a function of the frequency
        $\omega$ for different analytic continuation algorithms including,
        consistent constraints (solid blue), and three different variants of
        the maximum entropy method: historic (dashed yellow), classic (dash
        dotted green), and Bryan's method (dotted red) [see \cite{Levy2017}].
        Data is shown for $L = \beta = 64$ and $W =6$. The average is taken
        over all $384$ disordered samples.
    }
\end{figure}

\begin{figure*}
    \centering
    \includegraphics[width=\textwidth]{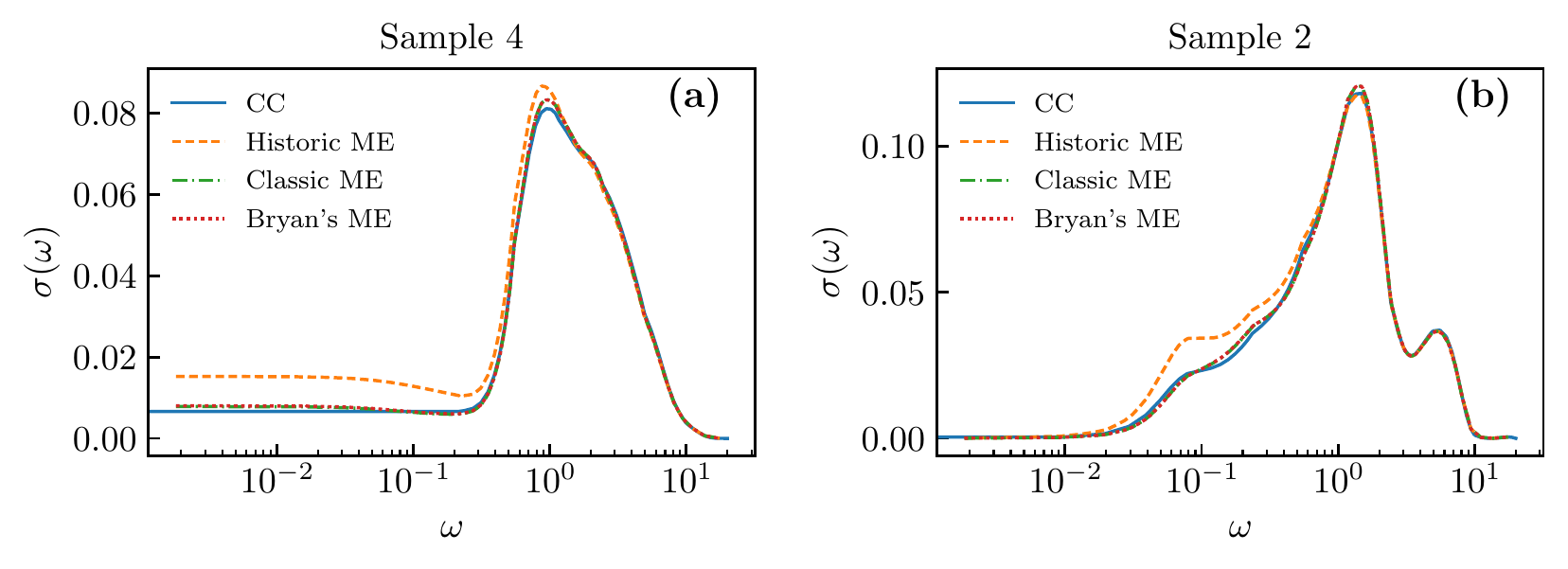}
    \caption{
        \label{fig:maxent_single}
        Comparison of different analytic continuation algorithms for single
        disorder realizations for the optical conductivity $\sigma(\omega)$ in
        a system with $L = 64$ and $\beta = 64$.  Each panel corresponds to
        different disorder realizations and different lines to different
        algorithms including, consistent constraints (solid blue), and three
        different maxent variants: historic (dashed yellow), classic (dash
        dotted green), and Bryan's method (dotted red) [see \cite{Levy2017}].
    }
\end{figure*}

\paragraph{Analytic continuation}
Here, we are interested in computing the optical conductivity $\sigma(\omega)$,
an observable that can be measured experimentally but not readily accessible by
numerical techniques.  By the dissipation-fluctuation theorem
\begin{equation}
    \chi(\imath\omega_n) = -\frac{2}{\pi} \int_0^{\infty}
    \frac{\omega^2}{\omega_n^2 + \omega^2} \sigma(\omega) \dd{\omega}.
\end{equation}
Finding $\sigma(\omega)$ is thus a standard ill conditioned inverse problem
when small fluctuations of the input due to statistical noise in the Monte
Carlo sampling lead to large fluctuations in the output results.  To solve this
problem we use a method of consistent constraints
\cite{Prokofev2013a,Goulko2017}. It allows us to restore the spectral density
$\sigma(\omega)$ from the corresponding correlation function
$\chi(\imath\omega_n)$.

As a consistency check we compare our results for the analytic continuation of
our data with a standard implementation \cite{Levy2017} of the maximum entropy
method \cite{Jarrell1996}. We note that maxent suffers from numerical
instabilities due to small errors on our data in Matsubara frequency domain and
it is able to find acceptable solutions only when artificially increasing the
errors and using the solutions found with the method of consistent constraints
as ``default model''.  The comparison is shown in \cref{fig:maxent_avg} in the
case of the disorder-averaged conductivity and in \cref{fig:maxent_single} for
single disorder realizations. Here, three different solutions are shown for
maxent (ME), corresponding to the three different variations of the maximum
entropy method available in the implementation of Ref.\cite{Levy2017}
(historic, classic and Bryan's method).  We see that, with exception of the
historic variant, all the solutions are essentially identical to each other and
our solution is accepted by maxent with little or no modifications.

\end{document}